\documentclass[10pt]{article}
\usepackage{amsmath,amssymb,enumerate} 

\usepackage{amsthm}

\usepackage{tikz}
\usetikzlibrary{arrows}
\usetikzlibrary{positioning}

\def\ZZ{\mathbb{Z}}
\def\ZZ1{\mathbf{Z}_{\geq 1}}

\title{On Nash-solvability of finite 
$n$-person deterministic graphical games; Catch 22 
}
\author{
Vladimir Gurvich\thanks{
HSE University, Moscow Russia; e-mail:
vgurvich@hse.ru ; vladimir.gurvich@gmail.com}
}

\begin{document}

\maketitle

\begin{abstract}
We consider finite $n$-person deterministic graphical (DG) games.
These games are modelled by finite directed graphs  
(digraphs) $G$  which may have directed cycles and, hence, infinite plays. 
Yet, it is assumed that all these plays are equivalent and 
form a single outcome  $c$, 
while the terminal vertices $V_T = \{a_1, \ldots, a_p\}$  form  $p$ remaining outcomes.    
\newline 
We study the existence of Nash equilibria (NE) in pure stationary strategies. 
It is known that NE exist when  $n=2$  and  may fail to exist when  $n > 2$.
Yet, the question becomes open for $n > 2$  under the following extra condition:   
\newline 
(C) For each of  $n$  players  
$c$  is worse than each of  $p$ terminal outcomes. 
\newline  
In other words, all players are interested in terminating the play, 
which is a natural assumption.
Moreover, Nash-solvability remains open 
even if we replace (C) by a weaker condition: 
\newline 
(C22) There exist no two players for whom  $c$  is better than 
(at least) two terminal outcomes.
\newline
We conjecture that such two players exist in each NE-free DG game, 
or in other words, that (C22)  implies Nash-solvability, for all $n$.  
\newline 
Recently, the DG games were extended to a wider class of the DG 
multi-stage (DGMS) games, whose outcomes are the  
strongly connected components (SCC) of digraph  $G$. 
Merging all outcomes of a DGMS game that correspond 
to its non-terminal SCCs we obtain a DG game. 
Clearly, this  operation respects Nash-solvability (NS).  
Basic conditions and conjectures related to NS 
can be extended from the DG to DGMS games:
in both cases NE exist if  $n=2$  and may  fail to exist when $n > 2$;  
furthermore,, we modify conditions (C) and (C22) 
to adapt them for the DGMS games.
\end{abstract}

\section{Deterministic graphical finite $n$-person games}
\label{s1}
Nash-solvability (NS) of these games is discussed in \cite{BGMOV18}. 
Here we use the concepts, definitions, and notation from this paper.   

NS holds for $n=2$ and may fail for  $n > 2$; see respectively  Sections 5 and 2 of \cite{BGMOV18}.  
The first NE-free example was constructed for $n=4$ in the preprint \cite{GO14A} 
and published in \cite{Gur15}  
\footnote{Vladimir Oudalov refused to co-author paper~\cite{Gur15} published in Russia.}  
Then, a much more compact example for  $n=3$  appeared in \cite[Section 3]{BGMOV18}.
In this example  $c$  is  the  worst outcome for  player 2 
but  for players 1 and 3  $c$  is better than (at least) two terminal outcomes. 
Somewhat surprisingly, computations show that such two players exist 
in each NE-free $n$-person DG game (for all  $n > 2$).  
In other words, condition (C22) given in Abstract implies NS.  
We call this conjecture "Catch 22". 
It is stronger than conjecture "(C) implies NS",   
since condition (C22) is weaker than (C).

\smallskip 

The latter means that  
every player strictly prefers each terminal play to each infinite one, 
which looks natural and can be easily interpreted  \cite[Section 3.2]{GN21A} for examples. 
If (C) does not imply NS  then there exists a DG game that has a cyclic NE 
(with outcome  $c$) but has no terminal NE 
(with an outcome $a \in V_T$) although 
$V_T$  can be reached from the initial position  \cite[Proposition 3]{BGMOV18}. 
This property looks strange but not impossible.

\medskip

To formulate (C22) more accurately
denote by  $k_c(i)$  the number of terminals that are
worse than  $c$  for player   $i \in I = \{1, \ldots, n\}$.
Then,  (C22)  does not hold if and only if
$k_c(i) \geq 2$  for (at  least) two players.

\bigskip

Digraph  $G$  is called {\em bidirected} if 
each non-terminal  move in it is reversible, that is,  
$v' v'' \in E(G)$  if and only if  $v'' v' \in E(G)$ 
unless  $v'$ or $v''$  is a terminal position. 
We conjecture that every $n$-person DG game on a finite bidirected digraph is NS.      

\section{Deterministic graphical multi-stage (DGMS) finite $n$-person games}
\label{s2}
Recently, DG games were generalized and DGMS games were introduced  
in  \cite[Section 1]{Gur18}; see also \cite[Section 4.1]{GN21B}. 
Here we use to the concepts, definitions, and notation from these papers.

It is well-known that every digraph $G$  is uniquely partitioned
into the strongly connected components (SCC).
An SCC will be called; 
\begin{itemize}
\item [] {\em terminal} if it has no outgoing edges (exits);
\item [] {\em interior} if it is not terminal and contains a directed cycle;
\item [] {\em transient} if it is neither terminal not interior.
\end{itemize}

Obviously, each transient component consists of a single non-terminal vertex.
Contracting every SCC to a single  vertex, one gets an acyclic digraph. 
Conversely, given an acyclic digraph  $G$, 
let us substitute some of its non-terminal vertices 
by strongly connected digraphs that contain directed cycles. 
They will form the interior SCCs of the obtained digraph  $G'$, 
while each of the remaining non-terminal vertices of  $G$  will form 
a transient SCC in   $G'$. 

\medskip

It seems logical to assign a separate outcome  $a$
(resp., $c$) to each terminal (resp., interior) SCC, thus,
getting the set of outcomes
\newline
$O = \{a_1, \ldots, a_p\} \cup \{c_1 \ldots, c_q\}$, 
which define a DGMS game. 
Merging the latter  $q$  outcomes in it  we obtain a DG game. 
Obviously, merging outcomes of a game form respects NS. 
Hence, an $n$-person DGMS game may have no NE when $n > 2$. 
The main examples of \cite{GO14A,BGMOV18} show this.  

\smallskip  

Yet, any two-person DGMS game has a NE, which 
can be determined by a slightly modified  backward induction (BI) procedure 
\cite[Section 1.5]{Gur18}; see also \cite[Section 4.1]{MN21B}. 
It was proven that finite $n$-person DG and DGMS game forms are tight, for all $n$. 
(Note that merging outcomes respects tightness.) 
Yet, tightness is equivalent with NS only when $n=2$; 
for $n > 2$  tightness is neither necessary nor sufficient for NS. 
Note also that the modified BI works only for  $n=2$  and 
even in this case the obtained NE may be not subgame perfect, 
unlike the NE obtained by the standard BI. 

\medskip 

Obviously, contracting each terminal SCC of a DGSM game 
to a single vertex one does not change the game.     
Finally, note that a DGSM game with a unique interior SCC is DG game. 

\bigskip 

Conditions (C) and (C22)  can be adapted to DGMS games as follows. 

\medskip 

$(C')$ For any player  $i \in I$ 
every interior SCC  $c$  is better than each terminal SCC  $a$. 

\medskip  

To formulate $(C'22)$  
denote by  $k(i, c_j)$  the number of terminal SCC that are
worse than  $c_j$  for player  $i \in I$.
Then,  (C'22)  does not hold if and only if
there are (at least) two distinct players  $i', i'' \in I$
and two, not necessarily distinct,
interior SCC $c_{j'}$ and $c_{j'}$  such that
$k(i', c_{j'}) \geq 2$   and   $k(i'', c_{j''}) \geq 2$.

Obviously $(C')$  implies   $(C'22)$. 
We conjecture that  $(C'22)$  implies NS of finite  $n$-person DGMS games 
(for  $n > 2$). .  

\section{Subgame perfect NE-free DG games} 
\label{s3} 
NE $s  = (s_1, \ldots, s_n)$  in a finite $n$-person DG game 
is called {\em uniform} if it is a NE with respect to 
every initial position $v \in V \setminus V_T$. 
In the literature uniform NE (UNE) are frequently 
referred to as {\em subgame perfect NE}. 
By definition, any UNE is a NE, but not vice versa. 
A large family of $n$-person UNE-free DG games satisfying (C) 
can be found in \cite[Section 3.3]{GN21A} for  $n > 2$,  
and even for  $n=2$  in \cite[the last examples in Figures 1 and 3]{BEGM12}.  

Every NE-free game contains a UNE-free subgame  \cite[Remark 3]{BGMOV18}. 
Indeed, consider an arbitrary NE-free DG game $\Gamma$ 
and  eliminate the initial position  $v_0$  from its  graph  $G$. 
The  obtained subgame $\Gamma'$ is UNE-free.    
Indeeed, assume for contradiction that  $\Gamma'$  has a UNE  $s  = (s_1, \ldots, s_n)$. 
Then, $\Gamma$ has a NE, which can be obtained by backward induction. 
The player beginning in  $v_0$  chooses a move that maximizes his/her reward,  
assuming that  $s$  is played in  $\Gamma'$  by all players. 
Clearly  $s$  extended by this move forms a NE in $\Gamma$, which is a contradiction. 
 
Thus, searching for a NE-free DG game 
(satisfying (C) or (C22))  
one should begin with a UNE-free DG game 
trying to extend it with an acyclic prefix. 
This was successfully reallized in  \cite{GO14A,BGMOV18}, 
where condition  (C22) (not to mention (C)) was waved. 
However, under these assumptions all tries failed.    

\subsection*{Acknowledgement}
The authors was partially supported by the RSF grant 20-11-20203.


\begin{thebibliography}{99}

\bibitem{BEGM12} 
Endre Boros, Khaled Elbassioni, Vladimir Gurvich, and Kazuhisa Makino, 
On Nash equilibria and improvement cycles in
pure positional strategies for Chess-like and Backgammon-like n-person games, 
Discrete Math. 312:4  (2012) 772--788.

\bibitem{BGMOV18}
Endre Boros, Vladimir Gurvich, Martin Milanic, Vladimir Oudalov, and Jernej Vicic, 
A three-person deterministic graphical game without Nash equilibria, 
Discrete Applied Math. 243 (2018) 21--38; 
https://arxiv.org/abs/1610.07701 . 


\bibitem{Gur15} 
Vladimir Gurvich,
A four-person chess-like game without Nash equilibria in pure stationary strategies,
Business Informatics 1:31 (2015) 68--76.

\bibitem{Gur18} 
Vladimir Gurvich, Backward induction in presence of cycles, 
Oxford Journal of Logic and Computation 28:7 (2018) 1635--1646.

\bibitem{GN21A} 
Vladimir Gurvich and Mariya Naumova,
On Nash-solvability of n-person graphical games under Markov's and a priori realizations;
https://arxiv.org/abs/2104.07542 

\bibitem{GN21B} 
Vladimir Gurvich and Mariya Naumova,  
Polynomial algorithms computing two lexicographically safe Nash equilibria
in finite two-person games with tight game forms given by oracles;
https://arxiv.org/abs/2108.05469 

\bibitem{GO14A}
Vladimir Gurvich and Vladimir Oudalov, 
A four-person chess-like game
without Nash equilibria in pure stationary strategies,
https://arxiv.org/abs/1411.0349 [math.CO], 2014.

%
%



\end{thebibliography}
\end{document}